\documentclass[amsmath,amssymb,showkeys,superscriptaddress,floatfix,prd,10pt,aps,twocolumn]{revtex4-2}
\usepackage{graphicx,epstopdf}
\usepackage{textcomp}
\usepackage[caption=false]{subfig}
\usepackage{multirow}
\usepackage{appendix}
\def\pslash{p\!\!\!\slash }
\def\qslash{q\!\!\!\slash }
\def\xslash{x\!\!\!\slash }
\def\eslash{\varepsilon\!\!\!\slash }

\def\vel{\left|}
\def\ver{\right|}
\usepackage{colortbl}
\usepackage{xcolor}
\usepackage[colorlinks=true, urlcolor=blue, linkcolor=green, citecolor=green]{hyperref}

\begin{document}

\title{Investigation of magnetic moment of $P_{cs}(4338)$ and $P_{cs}(4459)$ pentaquark states}

\author{Ula\c{s} \"{O}zdem}%
\email[]{ulasozdem@aydin.edu.tr}
\affiliation{Health Services Vocational School of Higher Education, Istanbul Aydin University, Sefakoy-Kucukcekmece, 34295 Istanbul, T\"{u}rkiye}

\date{\today}
 
\begin{abstract}
In this study, we discuss the magnetic moment of $P_{cs}(4338)$ and $P_{cs}(4459)$ hidden-charmed pentaquarks, which are closely related to their substructures. The magnetic moments of these states are calculated with the help of the QCD light-cone sum rules method with quantum numbers $I(J^P) = 0(1/2^-)$ and $I(J^P) = 0(3/2^-)$ for $P_{cs}(4338)$ and $P_{cs}(4459)$, respectively. Our predictions for the magnetic moment $\mu_{P_{cs}} = 0.34 \pm 0.12~\mu_N$ for the $P_{cs}(4338)$ state, and $\mu_{P_{cs}} = -1.67 \pm 0.58~\mu_N $ for the $P_{cs}(4459)$ state. 
As a by-product, the magnetic moments of the isospin$-1$ partners of these states have also been obtained. The magnetic moments are obtained as  $\mu_{P_{cs}} = 0.63 \pm 0.21~\mu_N$  and $\mu_{P_{cs}} = -3.33 \pm 1.04 ~\mu_N $ for the isospin-1 partners of the $P_{cs}(4338)$ and $P_{cs}(4459)$ states, respectively. Our results regarding the magnetic moments of these color singlet-singlet type pentaquark states are compared with the results in the literature.  
\end{abstract}

\maketitle

\section{Introduction}\label{motivation}

The discovery and investigation of hadron states with quantum numbers that go beyond the traditional hadron structure are one of the most attractive platforms of hadron spectroscopy due to the insight that these states can ensure the information on confinement, gluonic degrees of freedom and, on QCD in the GeV scale.  
Since the discovery of the X(3872) in 2003 by Belle Collaboration, many unconventional states were discovered in the past two decades.  The different elucidations have also appeared together with conventional hadrons, loosely bound molecules, compact tetraquarks or pentaquarks, kinematic effects, glueballs, hybrids, and so on.  Though there is still a long way to precisely figure out how the strong interaction binds these quarks and antiquarks together, this subject has become one of the most attractive research subjects in hadron physics.
For the recent experimental and theoretical progress on the unconventional states see, for instance, Refs.~\cite{Esposito:2014rxa,Esposito:2016noz,Olsen:2017bmm,Lebed:2016hpi,Nielsen:2009uh,Brambilla:2019esw,Agaev:2020zad,Chen:2016qju,Ali:2017jda,Guo:2017jvc,Liu:2019zoy,Yang:2020atz,Dong:2021juy,Dong:2021bvy,Meng:2022ozq, Chen:2022asf}.

In 2015, two pentaquark states, $P_{c }(4380)$ and $P_{c }(4450)$, were observed by the LHCb collaboration in the  $J/\psi\,p$ invariant mass distribution~\cite{Aaij:2015tga}. The LHCb Collaboration then updated its previous analysis and announced that it had observed three new pentaquark states, $P_{c}(4312)$, $P_{c}(4440)$, and $P_{c}(4457)$~\cite{Aaij:2019vzc}.  In 2020, the LHCb Collaboration announced a pentaquark state, $P_{cs}(4459)$, in the invariant mass spectrum of $J/\psi\Lambda$ in the $\Xi_b^0 \rightarrow J/\psi\,\Lambda\,K^-$ decay~\cite{Aaij:2020gdg}. The measured mass and width are $4458.8 \pm 2.7 ^{+4.7}_{-1.1}$ MeV and $17.3 \pm 6.5^{+8.0}_{-5.7}$ MeV respectively.  Since this pentaquark state is observed in the invariant mass distribution $J/\psi\Lambda$, the quark content is assumed to be $udsc\bar c$; However, the spin-parity quantum numbers have not yet been clarified.  In 2021, evidence for a new pentaquark structure, $P_c(4337)$,  was found by the LHCb Collaboration in $B_s \rightarrow J/\psi\, p \bar p$ decays \cite{LHCb:2021chn} and spin-parity quantum numbers of this state were predicted with $J^P = 1/2^+$ \cite{Shen:2017ayv}. Recently, the LHCb collaboration observed a new structure $P_{cs}(4338)$ in the $J/\psi\Lambda$ mass distribution in the $B^- \rightarrow J/\psi\Lambda^- p $ decays~\cite{Collaboration:2022boa}. The measured mass and width are $4338.2 \pm 0.7 \pm 0.4$ MeV and $7.0 \pm 1.2 \pm 1.3$ MeV respectively and the amplitude analysis prefers the $J^P = 1/2^-$ spin-parity quantum numbers and excludes the possibility of $J^P = 1/2^+$ at $90\%$ confidence level.  As regards its decay products, one can conclude that this newly reported state consists of $udsc\bar c$ quarks.  Before and after the experimental discoveries were announced, numerous studies were conducted to understand the  properties of  hidden-charm strange pentaquark states~
\cite{Shen:2020gpw, Wang:2019nvm, Wu:2010jy, Chen:2016ryt, Shen:2019evi, Xiao:2019gjd, Anisovich:2015zqa, Feijoo:2015kts, Lu:2016roh, Liu:2020hcv, Zou:2021sha, Karliner:2021xnq, Peng:2020hql, Zhu:2021lhd, Hu:2021nvs, Du:2021bgb,  Xiao:2021rgp, Chen:2020uif, Chen:2015moa, Chen:2016otp, Xiang:2017byz, Chen:2019bip, Chen:2020pac, Chen:2020opr, Chen:2020kco, Chen:2021tip, Chen:2022onm, Chen:2019asm, Wu:2021caw, Lu:2021irg, Yang:2021pio, Cheng:2021gca, Clymton:2021thh, Liu:2020ajv, Deng:2022vkv, Shi:2021wyt, Wang:2022gfb, Wang:2021itn, Wang:2020eep, Azizi:2021utt, Wang:2022neq,Karliner:2022erb, Wang:2022mxy,Yan:2022wuz,Meng:2022wgl,Burns:2022uha}.

Studies in the literature for currently experimentally observed pentaquark states indicate that spectroscopic analyzes may not be sufficient to detect the properties of such states.  In addition to the spectroscopic properties of these states, investigation of their strong, weak, and electromagnetic decays may contribute to the understanding of the nature of these states.  In the literature, there are few studies on the calculation of electromagnetic multipole moments and transitions magnetic moments of pentaquark states through different configurations~\cite{Wang:2016dzu, Ozdem:2018qeh, Ortiz-Pacheco:2018ccl,Xu:2020flp,Ozdem:2021btf, Li:2021ryu,Ozdem:2021ugy,Gao:2021hmv, Ozdem:2022iqk, Ozdem:2022vip}.   In this work, we study the magnetic moment of $P_{cs}(4338)$  and $P_{cs}(4459)$ pentaquark states (hereafter we will show these states as $P_{cs}^1$  and $P_{cs}^2$, respectively) making use of the QCD light-cone sum rule formalism in the color singlet-singlet structure.  The magnetic moments belong to the non-perturbative domain of QCD, and to achieve their calculations we need non-perturbative methods. The QCD sum rule is one of the effective techniques that consider the non-perturbative effects. One of the modifications of the traditional QCD sum rules technique is the QCD light-cone version where operator product expansion is applied over the twist of operators instead of the dimensions of operators~\cite{Chernyak:1990ag, Braun:1988qv, Balitsky:1989ry}.

This paper is organized in the following manner: In Sect. \ref{formalism}, we will briefly describe the formalism of the QCD light-cone sum rules method, which is used to obtain the magnetic moments of the $P_{cs}^1$ pentaquark in the first step and then the $P_{cs}^2$ pentaquark.
Sect. \ref{numerical} provides the details of numerical computations to get the numerical results.  Sect. \ref{summary} is reserved for summary and concluding remarks.
Explicit expressions of the magnetic moment of the $P_{cs}^1$ pentaquark state are presented in the Appendix.

%\begin{widetext}
 
\section{The QCD light-cone sum rules for the $P_{cs}$ pentaquark states}\label{formalism}

\subsection{Formalism of the $P_{cs}^1 $ state} 

The correlation function, which is essential for magnetic moment calculations, is written in the following form:
 \begin{eqnarray} \label{edmn01}
\Pi(p,q)&=&i\int d^4x e^{ip \cdot x} \langle0|T\left\{J^{P_{cs}^1}(x)\bar{J}^{P_{cs}^1}(0)\right\}|0\rangle _\gamma \, , 
\end{eqnarray}
where sub-indice $\gamma$ is the external electromagnetic field. Here, $J(x)$ represents the interpolating current of the state considered, which in our case $P_{cs}^1$ pentaquark state, and it is required for further calculations. This interpolating current is written with
isospin and spin-parity $I(J^P) = 0(1/2^-)$ as follows ~\cite{Wang:2022neq}:
\begin{align}\label{curpcs1}
J^{P_{cs}^1}(x)&=\frac{1}{\sqrt{2}}\Big \{ \mid \bar D^0 \Xi^0_c \rangle \, - \mid \bar D^- \Xi^+_c \rangle  \Big \}\nonumber\\
&=\frac{1}{\sqrt{2}} \Big \{ \big[\bar c^d(x)i \gamma_5 u^d(x)\big]\big[\varepsilon^{abc} d^{a^T}(x)C\gamma_5 s^b(x) \nonumber\\
& \times c^c(x)\big] 
- \big[\bar c^d(x)i \gamma_5 d^d(x)\big]\big[\varepsilon^{abc} u^{a^T}(x)C\gamma_5 s^b(x)
\nonumber\\
& \times c^c(x)\big] \Big\} \, , 
%%%%%%%%%%%%%%%%%%%%%%%%%%%%%%%%%%%%%%%%%%%%%%%%%%%%%%%%%
%%%%%%%%%%%%%%%%%%%%%%%%%%%%%%%%%%%%%%%%%%%%%%%%%%%%%%%%%
 \end{align}
where $a$, $b$, $c$ and  $d$ are color indices and the $C$ is the charge conjugation operator. 

We would like to point out that numerous possible interpolating currents can be written for pentaquark states, however, the number of possible interpolating currents that can be written can be slightly reduced when the QCD sum rules and the states to be investigated are considered.  Additionally, we should define the isospins of the interpolating currents to make reliable estimations, it is the key issue to solving the puzzle of those $P_{cs}$ states. Therefore, we construct the local color singlet-singlet type five-quark currents with the definite isospins, which couple potentially to the color singlet-singlet type $P_{cs}$ states rather than to the meson-baryon scattering states or thresholds.

To obtain the hadronic representation of the correlation function, we insert a full set of intermediate states $P_{c s}^1$ with the same quantum numbers as the interpolation currents into the correlation function. As a result, we get the following expression
 \begin{align}\label{edmn02}
\Pi^{Had}(p,q)&=\frac{\langle0\mid J^{P_{c s}^1}(x) \mid
{P_{c s}^1}(p, s) \rangle}{[p^{2}-m_{P_{c s}^1}^{2}]}\nonumber\\
& \times 
\langle {P_{c s}^1}(p, s)\mid
{P_{c s}^1}(p+q, s)\rangle_\gamma 
\nonumber\\
& \times 
\frac{\langle {P_{c s}^1}(p+q, s)\mid
\bar J^{P_{c s}^1}(0) \mid 0\rangle}{[(p+q)^{2}-m_{P_{c s}^1}^{2}]}+ \cdots 
\end{align}
%where  

The matrix elements $\langle0\mid J^{P_{c s}^1}\mid {P_{c s}^1}(p, s)\rangle$, $\langle {P_{c s}^1}(p+q, s)\mid\bar J^{P_{c s}^1}\mid 0\rangle $ and  $\langle
{P_{c s}^1}(p, s)\mid {P_{c s}^1}(p+q, s)\rangle_\gamma$ in Eq. (\ref{edmn02}) can be parameterized in terms of residue, spinor, and Lorentz invariant form factors as follows: 
%
%\begin{widetext}
\begin{align} 
\langle0\mid J^{P_{c s}^1}(x)\mid {P_{c s}^1}(p, s)\rangle=&\lambda_{P_{c s}^1} \gamma_5 \, u(p,s),\label{edmn04}\\
\nonumber\\
\langle {P_{c s}^1}(p+q, s)\mid\bar J^{P_{c s}^1}(0)\mid 0\rangle=&\lambda_{P_{c s}^1} \gamma_5 \, \bar u(p+q,s)\label{edmn004}
,\\
\nonumber\\
%\end{align}
%\begin{align}
\langle {P_{c s}^1}(p, s)\mid {P_{c s}^1}(p+q, s)\rangle_\gamma &=\varepsilon^\mu\,\bar u(p, s)\Big[\big[F_1(q^2)
\nonumber\\
&
+F_2(q^2)\big] \gamma_\mu +F_2(q^2)
\nonumber\\
& \times 
\frac{(2p+q)_\mu}{2 m_{P_{c s}^1}}\Big]\,u(p+q, s). \label{edmn005}
\end{align}

Then Eq. (\ref{edmn04}),  Eq. (\ref{edmn004})  and Eq. (\ref{edmn005}) are substituted in the Eq. (\ref{edmn02}) and some calculations are made, the following result is obtained for the hadronic side,
\begin{align}
\label{edmn05}
\Pi^{Had}(p,q)=&\lambda^2_{P_{c s}^1}\gamma_5 \frac{\Big(\pslash+m_{P_{c s}^1} \Big)}{[p^{2}-m_{{P_{c s}^1}}^{2}]}\varepsilon^\mu \Bigg[\big[F_1(q^2) %
%\nonumber\\
%&
+F_2(q^2)\big] \gamma_\mu
\nonumber\\
& 
+F_2(q^2)\, \frac{(2p+q)_\mu}{2 m_{P_{c s}^1}}\Bigg]  \gamma_5 
\frac{\Big(\pslash+\qslash+m_{P_{c s}^1}\Big)}{[(p+q)^{2}-m_{{P_{c s}^1}}^{2}]}. 
\end{align}
In obtaining the above expression, summation over spins of $P_{cs}^1$
\begin{align}
\label{edmn0004}
 \sum_s u(p,s)\bar u(p,s)&=\pslash+m_{P_{cs}^1},\\
  \sum_s u(p+q,s)\bar u(p+q,s)&=(\pslash+\qslash)+m_{P_{cs}^1},
\end{align}
have also been applied. 

The value of form factors $F_1(q^2)$ and $F_2(q^2)$ give us the  magnetic form factor $F_M(q^2)$ at different $q^2$ :
\begin{align}
\label{edmn07}
&F_M(q^2) = F_1(q^2) + F_2(q^2).
\end{align}
 For real photon, i.e. $q^2 = 0 $,  magnetic form factor $F_M (q^2 = 0)$ is proportional to the magnetic moment $\mu_{P_{c s}^1}$:
\begin{align}
\label{edmn08}
&\mu_{P_{c s}^1} = \frac{ e}{2\, m_{P_{c s}^1}} \,F_M(q^2 = 0).
\end{align}

The second representation of the correlation function, the QCD side, is achieved by employing the interpolating currents explicitly into the correlation function.  Then, the proper quark fields are contracted with the help of Wick's theorem and the demanded outcomes are achieved. As a result of the above procedures, the QCD representation of the correlation function is obtained as follows:
\begin{align}
\label{QCD1}
\Pi^{QCD}(p,q)&=- \frac{1}{2}\varepsilon^{abc} \varepsilon^{a^{\prime}b^{\prime}c^{\prime}}\, \int d^4x \, e^{ip\cdot x} \langle 0\mid \Big\{ 
\nonumber\\
& 
\, \mbox{Tr}\Big[\gamma_5 S_{u}^{dd^\prime}(x) \gamma_5  S_{c}^{d^\prime d}(-x)\Big]  
\mbox{Tr}\Big[\gamma_5 S_s^{bb^\prime}(x) \gamma_5 
\nonumber\\
& \times  \widetilde S_{d}^{aa^\prime}(x)\Big]
- \mbox{Tr}\Big[\gamma_5 S_{u}^{da^\prime}(x) \gamma_5  \widetilde S_s^{bb^\prime}(x) \gamma_5  
\nonumber\\
& \times S_{d}^{ad^\prime}(x)   \gamma_5 S_{c}^{d^\prime d}(-x)\Big] 
   - \mbox{Tr}\Big[\gamma_5 S_{d}^{da^\prime}(x) \gamma_5    \nonumber\\ 
&  \times \widetilde S_s^{bb^\prime}(x) \gamma_5  S_{u}^{ad^\prime}(x) \gamma_5 S_{c}^{d^\prime d}(-x)\Big]+
 \mbox{Tr}\Big[\gamma_5 
  \nonumber\\ 
& \times  S_{d}^{dd^\prime}(x) \gamma_5 S_{c}^{d^\prime d}(-x)\Big]  
\mbox{Tr}\Big[\gamma_5 S_s^{bb^\prime}(x)
  \gamma_5 
   \nonumber\\ 
& \times \widetilde S_{u}^{aa^\prime}(x)\Big] \Big\}
\mid 0 \rangle _\gamma \, , 
\end{align}
where   
%\begin{equation*}
$\widetilde{S}_{c(q)}^{ij}(x)=CS_{c(q)}^{ij\mathrm{T}}(x)C$ and,
%\end{equation*}%
 $S_{c}(x)$ and $S_{q}(x)$ are the  charm and light quark propagators, respectively. The expressions of
these propagators are given as~\cite{Yang:1993bp, Belyaev:1985wza}
\begin{align}
\label{edmn13}
S_{q}(x)&= \frac{1}{2 \pi x^2}\Big(i \frac{\xslash}{x^2}- \frac{m_q}{2}\Big) 
- \frac{\langle \bar qq \rangle }{12} \Big(1-i\frac{m_{q} \xslash}{4}   \Big)
\nonumber\\
&
- \frac{ \langle \bar qq \rangle }{192}
m_0^2 x^2  \Big(1-i\frac{m_{q} \xslash}{6}   \Big)
-\frac {i g_s }{32 \pi^2 x^2} ~G^{\mu \nu} (x) 
\nonumber\\
& \times 
\Big[\rlap/{x} 
\sigma_{\mu \nu} +  \sigma_{\mu \nu} \rlap/{x}
 \Big],%\\
%\nonumber\\
\end{align}%
%and
%%
\begin{align}
\label{edmn14}
S_{c}(x)&=\frac{m_{c}^{2}}{4 \pi^{2}} \Bigg[ \frac{K_{1}\Big(m_{c}\sqrt{-x^{2}}\Big) }{\sqrt{-x^{2}}}
+i\frac{{\xslash}~K_{2}\Big( m_{c}\sqrt{-x^{2}}\Big)}
{(\sqrt{-x^{2}})^{2}}\Bigg]
\nonumber\\
&
-\frac{g_{s}m_{c}}{16\pi ^{2}} \int_0^1 dv\, G^{\mu \nu }(vx)\Bigg[ (\sigma _{\mu \nu }{\xslash}
  +{\xslash}\sigma _{\mu \nu }) 
   \nonumber\\
  &
\times \frac{K_{1}\Big( m_{c}\sqrt{-x^{2}}\Big) }{\sqrt{-x^{2}}}
+2\sigma_{\mu \nu }K_{0}\Big( m_{c}\sqrt{-x^{2}}\Big)\Bigg],
\end{align}%
where $\langle \bar qq \rangle$, $G^{\mu\nu}$, $v$, and  $K_i$'s are light-quark condensate, the gluon field strength tensor, line variable, and modified Bessel functions of the second kind, respectively.   The first term of the light and heavy quark propagators correspond to the perturbative or free part and the rest belong to the interacting parts. 

The correlation function given in Eq. (\ref{QCD1}) receives both perturbative, i.e., when a photon interacts perturbatively with quark propagators, and nonperturbative, i.e., photon interacts with light quarks at a large distance, contributions. 
In the first part, the propagator of the quark interacting with the photon perturbatively is replaced by

\begin{align}
\label{free}
S^{free}(x) \rightarrow \int d^4y\, S^{free} (x-z)\,\rlap/{\!A}(z)\, S^{free} (z)\,,
\end{align}
and the remaining propagators in Eq.~(\ref{QCD1}) are substituted with the full quark propagators including the perturbative and nonperturbative parts.  Here we use $ A_\mu(z)=-\frac{1}{2}\, F_{\mu\nu}(z)\, z^\nu $ where  the electromagnetic field strength tensor is written as $ F_{\mu\nu}(z)=-i(\varepsilon_\mu q_\nu-\varepsilon_\nu q_\mu)\,e^{iq.z} $. The total perturbative contribution is achieved by performing the replacement mentioned above for the perturbatively interacting quark propagator with the photon and making use of the replacement of the remaining propagators by their free parts.

In the next part, one of the light quark propagators in Eq.~(\ref{edmn13}), defining the photon emission at large distances, is substituted by
\begin{align}
\label{edmn21}
S_{\alpha\beta}^{ab}(x) \rightarrow -\frac{1}{4} \big[\bar{q}^a(x) \Gamma_i q^b(x)\big]\big(\Gamma_i\big)_{\alpha\beta},
\end{align}
and the rest propagators are substituted with the full quark propagators.
 Here, $\Gamma_i$ represents the full set of Dirac matrices. Once 
Eq. (\ref{edmn21}) is plugged into Eq. (\ref{QCD1}), there appear matrix
elements of  $\langle \gamma(q)\vel \bar{q}(x) \Gamma_i q(0) \ver 0\rangle$
and $\langle \gamma(q)\vel \bar{q}(x) \Gamma_i G_{\alpha\beta}q(0) \ver 0\rangle$ kinds,
representing the nonperturbative contributions.  To calculate the nonperturbative contributions, we need these matrix elements which are parameterized in terms of photon wave functions with definite twists. The explicit expressions of the photon distribution amplitudes (DAs) are presented in Ref.~\cite{Ball:2002ps}. 
The QCD side of the correlation function can be obtained in terms of quark-gluon parameters 
as well as the DAs of the photon using Eqs.~(\ref{QCD1})$-$(\ref{edmn21}) and after performing the Fourier transformation to remove the calculations to the momentum space.

 As a final step, the $\eslash \qslash$ Lorentz structure is chosen from both representations and the coefficients of the Lorentz structure are matched from both the hadronic and QCD representations. To eliminate the effects of the continuum and higher states, Borel transformation and continuum subtraction are performed.  After these procedures, we obtain the demanded QCD light-cone sum rules for the magnetic moments:
\begin{align}
\label{edmn15}
\mu_{P_{c s}^1} \,\lambda^2_{P_{c s}^1}\, m_{P_{c s}^1} =e^{\frac{m^2_{P_{c s}^1}}{M^2}}\, \Delta_1^{QCD} (M^2,s_0).
\end{align}
The analytical expressions obtained for the $\Delta_1^{QCD} (M^2,s_0)$ function are given in Appendix \ref{appenda}.

\subsection{Formalism of the $P_{cs}^2 $ state} 

For the magnetic moment of $P_{cs}^2 $ state required correlation function is given as 
\begin{eqnarray} \label{Pc101}
\Pi_{\mu\nu}(p,q)&=&i\int d^4x e^{ip \cdot x} \langle0|T\left\{J_\mu^{P_{cs}^2}(x)\bar{J}_\nu^{P_{cs}^2}(0)\right\}|0\rangle _\gamma \, .
\end{eqnarray}
%where $J_\mu^{P_{cs}^2}(x)$ is interpolating current of  $P_{cs}^2 $ pentaquark state. 
The interpolating current used for $P_{cs}^2 $ pentaquark state with
isospin and spin-parity $I(J^P) = 0(3/2^-)$ is as follows~\cite{Wang:2022neq}:
\begin{align}\label{curpcs2}
J_\mu^{P_{cs}^2}(x)&=\frac{1}{\sqrt{2}}\Big \{ \mid \bar D^{*0} \Xi^0_c \rangle \, - \mid \bar D^{*-} \Xi^+_c \rangle  \Big \}
\nonumber\\
&=\frac{1}{\sqrt{2}} \Big \{ \big[\bar c^d(x) \gamma_\mu u^d(x)\big]\big[\varepsilon^{abc} d^{a^T}(x)C\gamma_5 s^b(x)\nonumber\\
& \times c^c(x)\big] 
- \big[\bar c^d(x)\gamma_\mu d^d(x)\big]\big[\varepsilon^{abc} u^{a^T}(x)C\gamma_5 s^b(x)
\nonumber\\
& \times c^c(x)\big] \Big\} \, . 
%%%%%%%%%%%%%%%%%%%%%%%%%%%%%%%%%%%%%%%%%%%%%%%%%%%%%%%%%
%%%%%%%%%%%%%%%%%%%%%%%%%%%%%%%%%%%%%%%%%%%%%%%%%%%%%%%%%
 \end{align}

The correlation function obtained depending on the hadron parameters is written as,
\begin{align}\label{Pc103}
\Pi^{Had}_{\mu\nu}(p,q)&=\frac{\langle0\mid  J_{\mu}^{P_{cs}^2}(x)\mid
{P_{cs}^2}(p,s)\rangle}{[p^{2}-m_{{P_{cs}^2}}^{2}]}
\nonumber\\
& \times  \langle {P_{cs}^2}(p,s)\mid
{P_{cs}^2}(p+q,s)\rangle_\gamma 
\nonumber\\
& \times 
\frac{\langle {P_{cs}^2}(p+q,s)\mid
\bar{J}_{\nu}^{P_{cs}^2}(0)\mid 0\rangle}{[(p+q)^{2}-m_{{P_{cs}^2}}^{2}]}+...
\end{align}
The matrix elements of the interpolating current 
between the vacuum and the $P_{cs}^2$ pentaquark are defined as
\begin{align}\label{lambdabey}
\langle0\mid J_{\mu}^{P_{cs}^2}(x)\mid {P_{cs}^2}(p,s)\rangle&=\lambda_{{P_{cs}^2}}u_{\mu}(p,s),\nonumber\\
%%%%%%%%%%%%%%%%%%%%%%%%%%%%%%%%%%
\langle {P_{cs}^2}(p+q,s)\mid
\bar{J}_{\nu}^{P_{cs}^2}(0)\mid 0\rangle &= \lambda_{{P_{cs}^2}}\bar u_{\nu}(p+q,s), 
\end{align}
where the $u_{\mu}(p,s)$, $u_{\nu}(p+q,s)$ and $\lambda_{{P_{cs}^2}}$ are the spinors and residue of the $P_{cs}^2$ pentaquark states, respectively.  
Summation over spins of $P_{cs}^2$ pentaquark state is performed as:
\begin{align}\label{raritabela}
\sum_{s}u_{\mu}(p,s)\bar u_{\nu}(p,s)&=-\Big(\pslash+m_{P_{cs}^2}\Big)\Big[g_{\mu\nu}
-\frac{1}{3}\gamma_{\mu}\gamma_{\nu}
\nonumber\\
& -\frac{2\,p_{\mu}p_{\nu}}
{3\,m^{2}_{{P_{cs}^2}}}+\frac{p_{\mu}\gamma_{\nu}-p_{\nu}\gamma_{\mu}}{3\,m_{{P_{cs}^2}}}\Big].
\end{align} 

The transition matrix element $\langle
{P_{cs}^2}(p)\mid {P_{cs}^2}(p+q)\rangle_\gamma$ entering Eq.
(\ref{Pc103}) can be written as follows
\cite{Weber:1978dh,Nozawa:1990gt,Pascalutsa:2006up,Ramalho:2009vc}:
\begin{align}\label{matelpar}
\langle {P_{cs}^2}(p,s)\mid {P_{cs}^2}(p+q,s)\rangle_\gamma &=-e\bar
u_{\mu}(p,s)\Bigg[F_{1}(q^2)g_{\mu\nu}\eslash 
\nonumber\\
& -
\frac{1}{2m_{{P_{cs}^2}}} 
\Big[F_{2}(q^2)g_{\mu\nu} 
\nonumber\\
&+F_{4}(q^2)\frac{q_{\mu}q_{\nu}}{(2m_{{P_{cs}^2}})^2}\Big]\eslash\qslash
\nonumber\\
&+
F_{3}(q^2)\frac{1}{(2m_{{P_{cs}^2}})^2}q_{\mu}q_{\nu}\eslash \Bigg]
\nonumber\\
& \times u_{\nu}(p+q,s).%\nonumber\\
\end{align}
where $F_i$'s are the Lorentz invariant form factors.

In principle, using the Eqs. (\ref{Pc103})$-$(\ref{matelpar}), we can get the final expression of the hadronic side of the correlation function, however in doing so we encounter two problems: not all Lorentz structures are independent, and the correlation function can also receive contributions from spin-1/2 particles that need to be eliminated. The matrix element 
of the current $J_{\mu}$ between spin-1/2 pentaquarks and vacuum is nonzero and is determined as
\begin{equation}\label{spin12}
\langle0\mid J_{\mu}(0)\mid B(p,s=1/2)\rangle=(A  p_{\mu}+B\gamma_{\mu})u(p,s=1/2).
\end{equation}
As is seen the unwanted spin-1/2 contributions are proportional to $\gamma_\mu$ and $p_\mu$.
 By multiplying both sides with $\gamma^\mu$ and using 
 the condition $\gamma^\mu J_\mu = 0$ one can determine the constant A in terms of B.
To remove the spin-1/2 pollutions and obtain only independent structures in the
correlation function, we apply the ordering for Dirac
matrices as $\gamma_{\mu}\pslash\eslash\qslash\gamma_{\nu}$ and eliminate terms 
with $\gamma_\mu$ at the beginning, $\gamma_\nu$ at the end and those proportional to $p_\mu$ and 
$p_\nu$~\cite{Belyaev:1982cd}. Consequently, employing Eqs. (\ref{Pc103})$-$(\ref{spin12})
the hadronic side of the $P_{cs}^2$ state takes the form,
\begin{align}\label{final phenpart}
\Pi^{Had}_{\mu\nu}(p,q)&=\frac{\lambda_{_{{P_{cs}^2}}}^{2}}{[(p+q)^{2}-m_{_{{P_{cs}^2}}}^{2}][p^{2}-m_{_{{P_{cs}^2}}}^{2}]} \nonumber\\
&\times
\Bigg[  g_{\mu\nu}\pslash\eslash\qslash \,F_{1}(q^2) 
%\nonumber\\
%&
-m_{{P_{cs}^2}}g_{\mu\nu}\eslash\qslash\,F_{2}(q^2)
\nonumber\\
&
-
\frac{F_{3}(q^2)}{4m_{{P_{cs}^2}}}q_{\mu}q_{\nu}\eslash\qslash%\,\nonumber\\
%&
-
\frac{F_{4}(q^2)}{4m_{{P_{cs}^2}}^3}(\varepsilon.p)q_{\mu}q_{\nu}\pslash\qslash 
\nonumber\\
&
+
\cdots %\mbox{other independent structures}
\Bigg].
\end{align}

The final expression of the hadronic representation for the chosen Lorentz structures is obtained as follows:

\begin{eqnarray}
\Pi^{Had}_{\mu\nu}(p,q)&=&\Pi_1^{Had}g_{\mu\nu}\pslash\eslash\qslash \,
+\Pi_2^{Had}g_{\mu\nu}\eslash\qslash\,+
...,
\end{eqnarray}
where $ \Pi_1^{Had} $ and $ \Pi_2^{Had} $ are functions of the form factors $ F_1(q^2) $ and  $ F_2(q^2) $, respectively; and other independent structures and form factors are denoted by dots.

The magnetic form factor, $G_{M}(q^2)$, is characterized with respect to the form factors $F_{i}(q^2)$ as follows~\cite{Weber:1978dh,Nozawa:1990gt,Pascalutsa:2006up,Ramalho:2009vc}:
\begin{align}
G_{M}(q^2) &= [ F_1(q^2) + F_2(q^2)] ( 1+ \frac{4}{5}
\tau ) -\frac{2}{5} [ F_3(q^2)  
\nonumber\\
&
+ 
F_4(q^2)] \tau ( 1 + \tau ), %\nonumber\\
\end{align}
  where $\tau
= -\frac{q^2}{4m^2_{{P_{cs}^2}}}$. At $q^2=0$, the magnetic moment
is obtained  with respect to the functions $F_1(q^2=0)$ and $F_2(q^2=0)$ form factors as:
\begin{eqnarray}\label{mqo1}
G_{M}(q^2=0)&=&F_{1}(q^2=0)+F_{2}(q^2=0).
\end{eqnarray}
The  magnetic moment of the spin-3/2 state, ($\mu_{{P_{cs}^2}}$), is described as follows,% in the following way:
 \begin{eqnarray}\label{mqo2}
\mu_{{P_{cs}^2}}&=&\frac{e}{2m_{{P_{cs}^2}}}G_{M}(q^2=0).
\end{eqnarray}

When the above procedures are applied, the analysis concerning hadronic parameters, which is the first step of the QCD light-cone sum rule analysis, is completed.

The next step in the QCD light-cone sum rule analysis is to obtain the correlation function with the quark-gluon parameters as well as the photon DAs. Doing again the steps in the previous subsection yields the following result:
\begin{align}
\Pi_{\mu\nu}^{QCD}(p,q)&= \frac{i}{2}\varepsilon^{abc} \varepsilon^{a^{\prime}b^{\prime}c^{\prime}}\, \int d^4x \, e^{ip\cdot x} \langle 0\mid \Big\{ 
\nonumber\\
&
\, \mbox{Tr}\Big[\gamma_\mu S_{u}^{dd^\prime}(x) \gamma_\nu  S_{c}^{d^\prime d}(-x)\Big]  
\mbox{Tr}\Big[\gamma_5 S_s^{bb^\prime}(x) \gamma_5 
\nonumber\\
& \times \widetilde S_{d}^{aa^\prime}(x)\Big]
- \mbox{Tr}\Big[\gamma_\mu S_{u}^{da^\prime}(x) \gamma_5  \widetilde S_s^{bb^\prime}(x) \gamma_5 
\nonumber\\
& \times S_{d}^{ad^\prime}(x) \gamma_\nu S_{c}^{d^\prime d}(-x)\Big] 
%   \nonumber\\ 
%& 
- \mbox{Tr}\Big[\gamma_\mu S_{d}^{da^\prime}(x) \gamma_5 \nonumber\\
& \times \widetilde S_s^{bb^\prime}(x) \gamma_5  S_{u}^{ad^\prime}(x) \gamma_\nu S_{c}^{d^\prime d}(-x)\Big]
  \nonumber\\ 
&+ \mbox{Tr}\Big[\gamma_\mu S_{d}^{dd^\prime}(x) \gamma_\nu  S_{c}^{d^\prime d}(-x)\Big]  
\mbox{Tr}\Big[\gamma_5 S_s^{bb^\prime}(x)
   \nonumber\\
& \times 
\gamma_5
\widetilde S_{u}^{aa^\prime}(x)\Big] \Big\}
\mid 0 \rangle _\gamma \, , 
\end{align}

As a result, the QCD representation of the correlation function concerning the chosen Lorentz structures is obtained as:

\begin{eqnarray}
\Pi^{QCD}_{\mu\nu}(p,q)&=&\Pi_{1}^{QCD}g_{\mu\nu}\pslash\eslash\qslash \,
+\Pi_{2}^{QCD}g_{\mu\nu}\eslash\qslash\,+
....
\end{eqnarray}
where $ \Pi_1^{QCD} $  and $ \Pi_2^{QCD} $ are functions of the QCD degrees of freedom and photon DAs parameters. 
The analytical expressions obtained for $ \Pi_i^{QCD} $  functions are not given in the text because they are very lengthy.

The correlation function has been obtained in connection with both hadronic and quark-gluon parameters. For magnetic moment analysis, the QCD and hadronic representations of the correlation function are equated using the quark-hadron duality approach.
For the form factor $F_1$ and $F_2$, we obtain the QCD light-cone sum rules by matching the coefficients of the Lorentz structures $g_{\mu\nu}\pslash\eslash\qslash$ and $g_{\mu\nu}\eslash\qslash$, respectively.
\begin{eqnarray}
\Pi^{Had}_{\mu\nu}(p,q)= \Pi^{QCD}_{\mu\nu}(p,q).
\end{eqnarray}

Analytical results are obtained for both $P_{cs}^1$ and $P_{cs}^2$ color singlet-singlet type pentaquark states. The next step would be to perform numerical calculations for these pentaquark states.  
We should also mention that the magnetic moments of the isospin-1 partners of the $P_{cs}^1$ and $P_{cs}^2$  states have also been calculated. To do this, we substitute the $^{\backprime\backprime}+^{\prime\prime}$ sign instead of the $^{\backprime\backprime}-^{\prime\prime}$ sign in Eqs. (\ref{curpcs1}) and (\ref{curpcs2}). For the sake of brevity, only the analytical results of the isospin-0  $P_{cs}^1$ and $P_{cs}^2$  states have been presented.

%\end{widetext}

At the end of this section, we should remark that to derive Eqs. (\ref{edmn02}) and (\ref{Pc103}), we assume that the physical side of the QCD sum rule can be approximated by a single pole approximation. However, in the case of the multiquark states Eqs. (\ref{edmn02}) and (\ref{Pc103}) receives contributions from two-hadron reducible terms as well. Two-hadron contaminating terms have to be considered when extracting parameters of multiquark states ~\cite{Kondo:2004cr,Lee:2004xk}. For the multiquark systems, they lead to modification in the quark propagator
\begin{equation}
\frac{1}{m^{2}-p^{2}}\rightarrow \frac{1}{m^{2}-p^{2}-i\sqrt{p^{2}}\Gamma (p)%
},  \label{eq:Modif}
\end{equation}%
where $\Gamma (p)$ is the finite width of the multiquark systems generated by two-hadron scattering states.  When these effects are properly considered in the QCD sum rules,  they rescale the residue of the multiquark states under consideration leaving its mass unchanged. Detailed calculations show that the two-hadron contributions are small, and they can be neglected  (see Refs. \cite{Albuquerque:2021tqd,Albuquerque:2020hio,Wang:2015nwa,Agaev:2018vag,Sundu:2018nxt,Wang:2019hyc,Wang:2020iqt,Wang:2019igl,Wang:2020cme}). Therefore, aforementioned contributions are negligible, and it is enough to employ the zero-width single-pole approximation.

\section{Numerical analysis and discussion}\label{numerical}

The QCD light-cone sum rule for the magnetic moment of the $P_{cs}$ color singlet-singlet type pentaquark states contains many input parameters whose numerical values we need. 
The values of these parameters we use in the analysis are given as follows: $m_u=m_d=0$, $m_s =93.4^{+8.6}_{-3.4}\,\mbox{MeV}$, $m_c = 1.27 \pm 0.02\,\mbox{GeV}$~\cite{Workman:2022ynf}, $m_{P_{cs}^1} = 4338.2 \pm 0.7 \pm 0.4$ MeV,   $m_{P_{cs}^2} = 4458.8 \pm 2.9^{+4.7}_{-1.1}$ MeV,   $f_{3\gamma}=-0.0039~\mbox{GeV}^2$~\cite{Ball:2002ps},  
%$m_b = (4.78\pm 0.06)\,GeV$, 
 $\langle \bar ss\rangle = 0.8\, \langle \bar uu\rangle$ $\,\mbox{GeV}^3$ with $\langle \bar uu\rangle = 
\langle \bar dd\rangle=(-0.24 \pm 0.01)^3\,\mbox{GeV}^3$ \cite{Ioffe:2005ym},
$m_0^{2} = 0.8 \pm 0.1 \,\mbox{GeV}^2$ \cite{Ioffe:2005ym}, 
$\langle g_s^2G^2\rangle = 0.88~ \mbox{GeV}^4$~\cite{Nielsen:2009uh},  $\lambda_{P_{cs}^1} = (1.43^{+0.19}_{-0.18}) \times 10^{-3}$~GeV$^6$ and $\lambda_{P_{cs}^2} = (1.55^{+0.20}_{-0.19}) \times 10^{-3}$~GeV$^6$~\cite{Wang:2022neq}.  %$\chi=-2.85 \pm 0.5~\mbox{GeV}^{-2}$~\cite{Rohrwild:2007yt},. 
The photon DAs and explicit expressions of the wave functions used in these DAs and their input parameters are borrowed from Ref.~\cite{Ball:2002ps}.

 The QCD light-cone sum rules for magnetic moments of these pentaquarks are obtained as a function of the Borel mass parameter $M^2$ and continuum threshold $s_0$. To get reliable QCD light-cone sum rules to result, one should define appropriate working regions for these two helping parameters. The operator product expansion (OPE)  convergence and pole dominance (PC) limitations are widely used to define the working regions of these two helping parameters. Bearing in mind these limitations, the following working regions are obtained for these two helping parameters as a result of the numerical analysis,
\begin{align}
 &5.0~\mbox{GeV}^2 \leq M^2 \leq 7.0~\mbox{GeV}^2, \nonumber\\
 & 22.5~\mbox{GeV}^2 \leq s_0 \leq 24.5~\mbox{GeV}^2,%~ \mbox{for the $P_{cs}^1$} 
\end{align}
for the $P_{cs}^1$ state and
\begin{align}
 &5.0~\mbox{GeV}^2 \leq M^2 \leq 7.0~\mbox{GeV}^2, \nonumber\\
 & 23.5~\mbox{GeV}^2 \leq s_0 \leq 25.5~\mbox{GeV}^2,%~ \mbox{for the $P_{cs}^2$} 
\end{align}
for the $P_{cs}^2$ state. 

\begin{figure}[t]
%\centering
\subfloat[]{\includegraphics[width=0.45\textwidth]{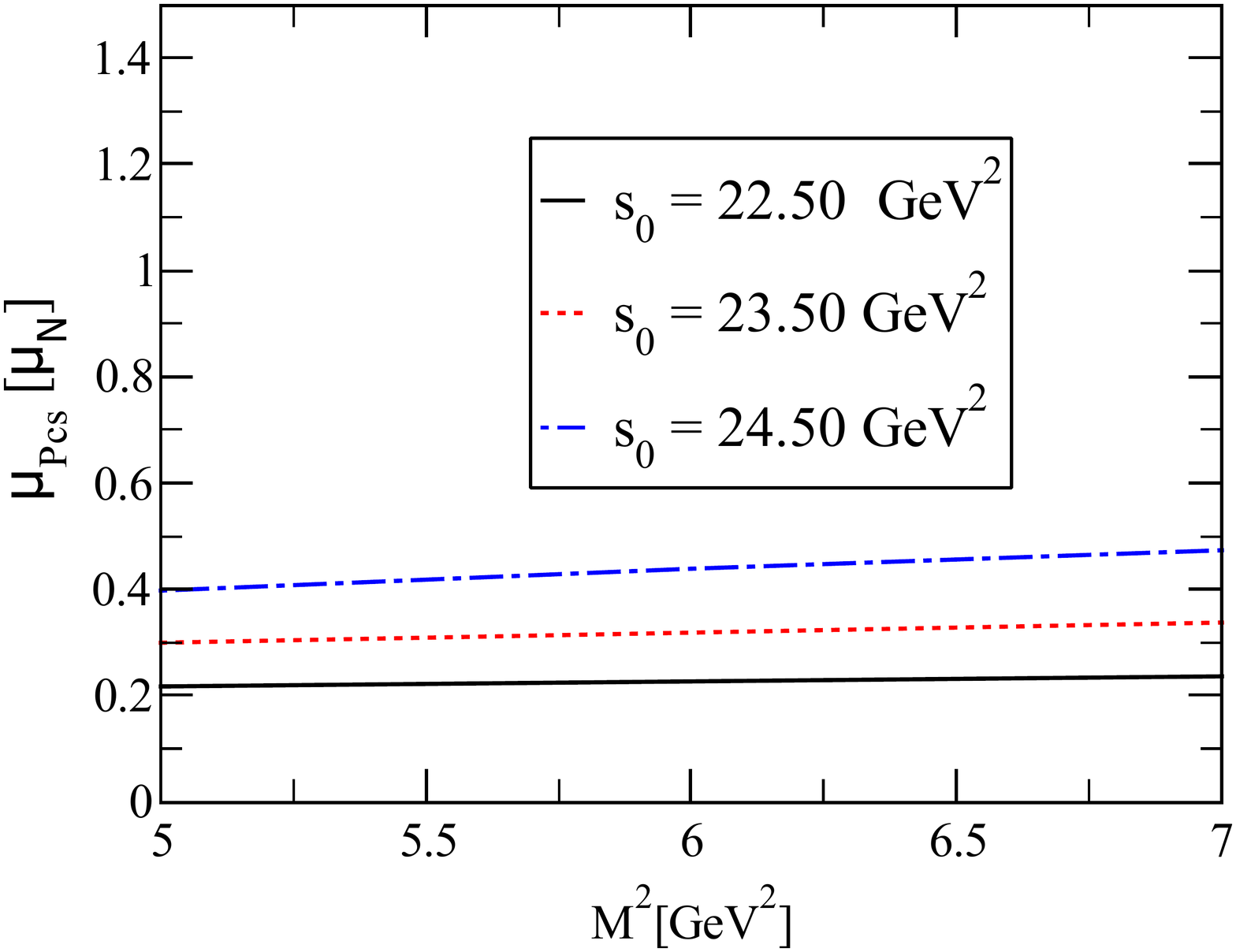}}\\
\subfloat[]{\includegraphics[width=0.45\textwidth]{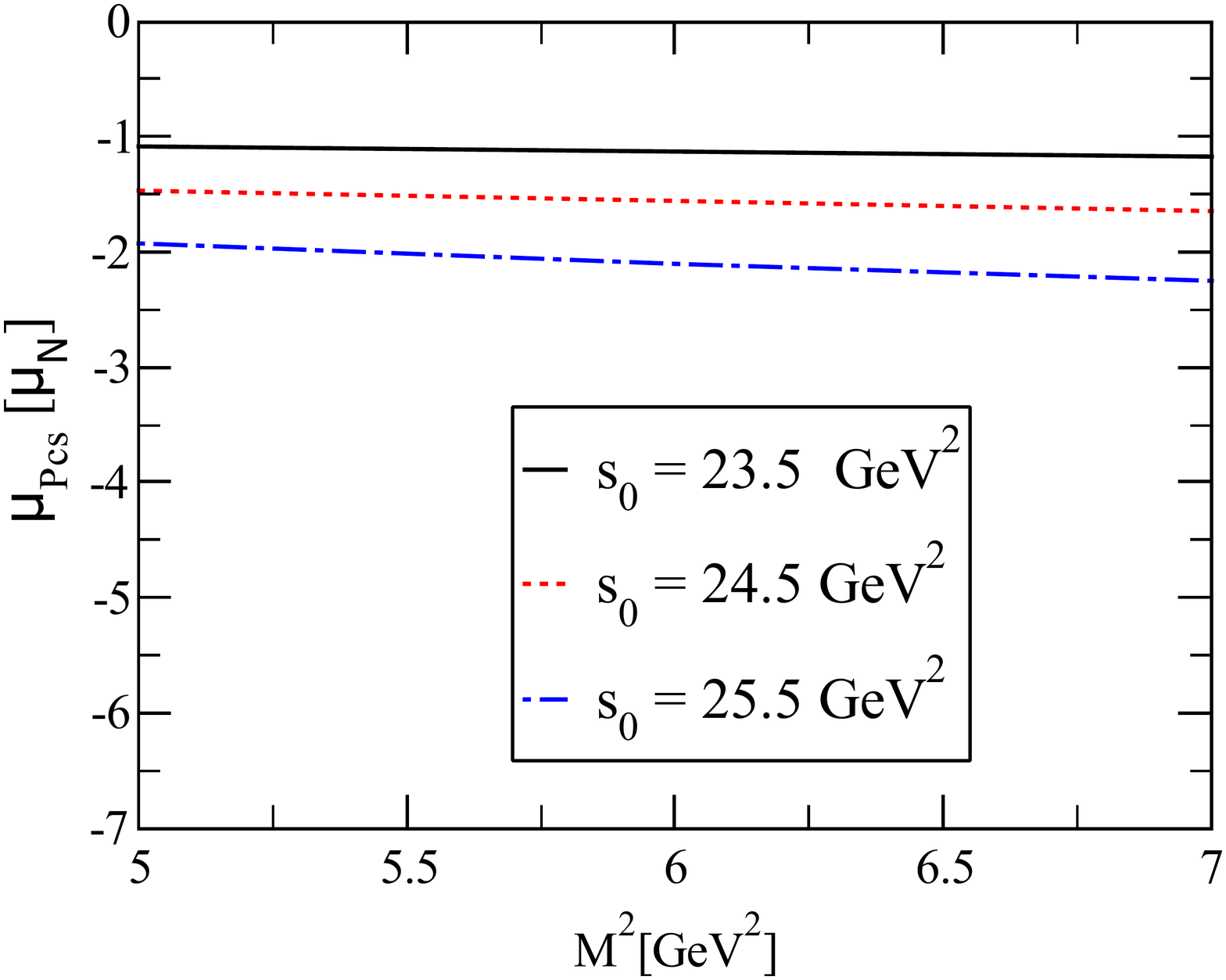}}%\\
 \caption{The magnetic moments of isospin-0 $([I,I_3] = [0,0])$ pentaquark states versus $M^2$ at three different values of $s_0$; (a) for $P_{cs}^{1}$ state and (b) for  $P_{cs}^{2}$ state (in unit of $\mu_N$).}
 \label{Msqfig1}
  \end{figure}

Using the above working regions for the $M^2$ and $s_0$, the PC changes in the intervals $60 \% \geq$ PC $\geq 33 \%$. At $M^2_{max} =7.0~$GeV$^2$, the PC is equal to $33\%$, whilst at $M^2_{min} =5.0~$GeV$^2$, it is equal to $60\%$. In the standard analysis of QCD sum rules, the PC should be larger than $50\%$ for baryons and mesons. In the case of multi-quark states, it turns out to be as PC $ > 20\%$. When we investigate the OPE convergence, we have obtained that the contribution of the higher dimensional terms in OPE is less than $\sim 2 \%$, thus the convergence of the sum rules is ensured. 
 As is seen the chosen working regions for $M^2$ and $s_0$ fulfill the requirements of the QCD light-cone sum rules technique. 
 For completeness, in Fig. \ref{Msqfig1}, we also depict the dependence of the magnetic moment of  $P_{cs}^1$ and $P_{cs}^2$ states, on the Borel mass parameter, $M^2$ at various values of $s_0$. As one can see from this figure, the variation of magnetic moments concerning $M^2$ is reasonably stable. Though the variation is high compared to $s_0$, this variation stays within the error limits of the QCD light-cone sum rules method.
 %%%%%%%%%%%%%%%%%%%%%%%%%%%%%%%%%%%%%%%%%%%%%

 % \begin{widetext}
 
%  \end{widetext}
 
 Now that we have determined the numerical values of all the input parameters, we can start performing the numerical analysis. Our final results for magnetic moments are given as:
  \begin{align}
  \mu_{P_{cs}^1} &= 0.34 \pm 0.12~\mu_N, \nonumber\\
  \mu_{P_{cs}^2} &= -1.67 \pm 0.58~\mu_N.
 \end{align}
 
The errors in the magnetic moment results given above are due to the input parameters, auxiliary parameters such as $s_0$ and $M^2$, as well as the input parameters used in the photon DAs.

In Refs.~\cite{Li:2021ryu, Gao:2021hmv}, the magnetic moments of the pentaquark states have been investigated employing the constituent quark model based on different combinations of the flavor-spin wave functions of the component hadrons. In Ref.~\cite{Gao:2021hmv}, the magnetic moment of  the $P_{cs}^1$ state  is obtained as $\mu_{P_{cs}^1} = 0.377~\mu_N$. For  the $P_{cs}^2$ state, while $\mu_{P_{cs}^2} = 0.465~\mu_N$ was found for magnetic moment in Ref.~\cite{Li:2021ryu}, $\mu_{P_{cs}^2} = - 0.231~\mu_N$ result was obtained in Ref.~\cite{Gao:2021hmv}. 
Our result for the magnetic moment of state $P_{cs}^1$ seems to be consistent with the result obtained in Ref.~\cite{Gao:2021hmv}. For the $P_{cs}^2$ pentaquark state, there are large discrepancies among results not only in the magnitude but also by the sign.

 Our final remark is that the magnetic moment of the  isospin-1 $([I, I_3] = [1,0])$ partners of these pentaquark states have also been obtained. The results are given as
 \begin{align}
  \mu_{P_{cs}^1} &= 0.63 \pm 0.21~\mu_N,  \nonumber\\
  \mu_{P_{cs}^2} &= -3.33 \pm 1.04~\mu_N.
 \end{align} 
When these results are examined, it is seen that the difference between the magnetic moment values of isospin-0, $([I,I_3]=[0,0])$, and isospin-1, $([I,I_3]=[1,0])$, pentaquark states is almost twice.
In Ref.~\cite{Gao:2021hmv}, the magnetic moments of the isospin-1 partners of these pentaquark states have been calculated and the results are obtained as $\mu_{P_{cs}^1} = 0.377~\mu_N$ and $\mu_{P_{cs}^2} = 0.465~\mu_N$. It is seen that the obtained results are not consistent with each other.
  
It appears from these results that different approaches lead to quite different estimates for the $P_{cs}^1$ and $P_{cs}^2$ magnetic moments that can be used to distinguish these models. It seems that more study is needed to understand the current situation. Hopefully, our study may attract the lattice QCD and experiment plans in the future.

\section{summary and concluding remarks}\label{summary}

Searching for multi-quark states is an interesting topic in hadron physics.
 The recently reported pentaquark state, the hidden-charmed strange $P_{cs}(4338)$, added a new member to the pentaquark family. The mass and width of this state were measured as  $4338.2 \pm 0.7 \pm 0.4$ MeV and $7.0 \pm 1.2 \pm 1.3$ MeV respectively.  
Inspired by this, in this study, we discuss the magnetic moment of $P_{cs}(4338)$ and $P_{cs}(4459)$ hidden-charmed pentaquarks, which are closely related to their substructures. The magnetic moments of these states are calculated with the help of the QCD light-cone sum rules method with quantum numbers $I(J^P) = 0(1/2^-)$ and $I(J^P) = 0(3/2^-)$. Our predictions for the magnetic moment $\mu_{P_{cs}} = 0.34 \pm 0.12~\mu_N$ for the $P_{cs}(4338)$ state and $\mu_{P_{cs}} = -1.67 \pm 0.58~\mu_N $ for the $P_{cs}(4459)$ state. 
As a by-product, the magnetic moments of the isospin$-1$ partners of these states have also been obtained. The magnetic moments are obtained as  $\mu_{P_{cs}} = 0.63 \pm 0.21~\mu_N$  and $\mu_{P_{cs}} = -3.33 \pm 1.04 ~\mu_N $ for the isospin$-1$ partners of the $P_{cs}(4338)$ and $P_{cs}(4459)$ states, respectively.  
Our results regarding the magnetic moments of these color singlet-singlet type pentaquark states are compared with the results in the literature.

 It is important to examine features of such multiquark states theoretically from different perspectives not only to ensure some insights into future experiments but also to better figure out the inner structure of these pentaquark states. The magnetic moments of hadronic states play an important role in understanding their internal structure.  We hope that our current research, together with existing studies in the literature, can inspire our colleagues to focus on the electromagnetic properties of pentaquark states in the future. With these efforts, our information on the features of pentaquark states will turn into more abundant. We are looking forward to future experimental progress along these directions.

%\section{Acknowledgments}

  \begin{widetext}
  \appendix
  \section*{Appendix: Explicit expression for \texorpdfstring{$\Delta_1^{QCD} (M^2,s_0)$}{}}\label{appenda}
 In this Appendix, we present the explicit expressions of the function $\Delta_1^{QCD} (M^2,s_0)$ for the magnetic moment of the $P_{cs}^1$ state entering into the sum rule.

 \begin{align}
  \Delta_1^{QCD} (M^2,s_0) &=\frac {m_c P_ 1 P_ 2 } {679477248 \pi^5}\Bigg[ 2 e_c \Bigg\{-6 m_c \Big (5 I[0, 3, 1, 0] + 2 I[0, 3, 1, 1] - 
       7 I[0, 3, 1, 2] - 2 (5 I[0, 3, 2, 0] + I[0, 3, 2, 1]) \nonumber\\
       &+ 
       5 I[0, 3, 3, 0]\Big) + 
    3 m_ 0^2 \Big (18 m_c (I[0, 2, 1, 0] - I[0, 2, 1, 1] - 
          I[0, 2, 2, 0]) - 
       m_s \big (I[0, 2, 1, 0] - 2 I[0, 2, 1, 1] 
       \nonumber\\
       &+ I[0, 2, 1, 2] - 
           2 I[0, 2, 2, 0] + 2 I[0, 2, 2, 1] + I[0, 2, 3, 0] - 
           2 I[1, 1, 1, 0] - 4 I[1, 1, 1, 1] + 2 I[1, 1, 1, 2] 
           \nonumber\\
       &- 
           4 I[1, 1, 2, 0] + 4 I[1, 1, 2, 1] + 
           2 I[1, 1, 3, 0]\big)\Big) - 
    4 m_s \Big (31 I[0, 3, 1, 0] - 63 I[0, 3, 1, 1] + 
        33 I[0, 3, 1, 2] 
        \nonumber\\
       &- I[0, 3, 1, 3] - 62 I[0, 3, 2, 0] + 
        64 I[0, 3, 2, 1] - 2 I[0, 3, 2, 2] + 31 I[0, 3, 3, 0] - 
        I[0, 3, 3, 1] + 3 I[1, 2, 1, 1] \nonumber\\
       &- 6 I[1, 2, 1, 2] + 
        3 I[1, 2, 1, 3] 
        - 6 I[1, 2, 2, 1] + 6 I[1, 2, 2, 2] + 
        6 I[1, 2, 3, 1]\Big)\Bigg\}
        \nonumber\\
      %  \end{align}
       % \begin{align}
        %%%%%%%%%%%%%%%%%%%%%%%%%%%%%%%%%%%%%%%%%%%%%%
        &-e_d \Bigg\{3 m_ 0^2 \Big (2  m_s \big (I[0, 2, 1, 0] - 
          2 I[0, 2, 1, 1] + I[0, 2, 1, 2] - 2 I[0, 2, 2, 0] + 
          2 I[0, 2, 2, 1] + I[0, 2, 3, 0] 
          \nonumber\\
       &- 2 I[1, 1, 1, 0] - 
          4 I[1, 1, 1, 1] + 2 I[1, 1, 1, 2] - 4 I[1, 1, 2, 0] + 
          4 I[1, 1, 2, 1] + 2 I[1, 1, 3, 0]\big) 
          \nonumber\\
       &+ 
       m_c \big (9 I[0, 2, 1, 0] - 18 I[0, 2, 1, 1] + 
           9 I[0, 2, 1, 2] - 14 I[0, 2, 2, 0] + 14 I[0, 2, 2, 1] + 
           5 I[0, 2, 3, 0] 
           \nonumber\\
       &- 10 I[1, 1, 1, 0] - 28 I[1, 1, 1, 1] + 
           18 I[1, 1, 1, 2] - 20 I[1, 1, 2, 0] + 28 I[1, 1, 2, 1] + 
           10 I[1, 1, 3, 0]\big)\Big) 
           \nonumber\\
       &+ 
    4 m_c \Big (5 I[0, 3, 1, 0] - 19 I[0, 3, 1, 1] + 
       21 I[0, 3, 1, 2] - 7 I[0, 3, 1, 3] - 10 I[0, 3, 2, 0] + 
       24 I[0, 3, 2, 1] 
       \nonumber\\
       &- 12 I[0, 3, 2, 2] + 5 I[0, 3, 3, 0] - 
       5 I[0, 3, 3, 1] + 15 I[1, 2, 1, 1] - 36 I[1, 2, 1, 2] + 
       21 I[1, 2, 1, 3] 
       \nonumber\\
       &- 30 I[1, 2, 2, 1] + 36 I[1, 2, 2, 2] + 
       15 I[1, 2, 3, 1]\Big) + 
    8  m_s \Big (I[0, 3, 1, 0] - 3 I[0, 3, 1, 1] + 3 I[0, 3, 1, 2] 
    \nonumber\\
       &- 
        I[0, 3, 1, 3] - 2 I[0, 3, 2, 0] + 4 I[0, 3, 2, 1] - 
        2 I[0, 3, 2, 2] + I[0, 3, 3, 0] - I[0, 3, 3, 1] + 
        3 I[1, 2, 1, 1] 
        \nonumber\\
       &- 6 I[1, 2, 1, 2] + 3 I[1, 2, 1, 3] 
        - 
        6 I[1, 2, 2, 1] + 6 I[1, 2, 2, 2] + 3 I[1, 2, 3, 1]\Big)\Bigg\}
        \nonumber%\\
        %%%%%%%%%%%%%%%%%%%%%%%%%%%%%%%%%%%%%%%%%%%%%%%%%%%%%%%%%
          \end{align}
       \begin{align}
        &
        -2 e_s m_c \Bigg\{3 m_ 0^2 \Big (I[0, 2, 1, 0] - 2 I[0, 2, 1, 1] + 
       I[0, 2, 1, 2] - 6 I[0, 2, 2, 0] + 6 I[0, 2, 2, 1] + 
       5 I[0, 2, 3, 0] 
       \nonumber\\
       &- 10 I[1, 1, 1, 0] - 12 I[1, 1, 1, 1] + 
       I[1, 1, 1, 2] - 20 I[1, 1, 2, 0] + 12 I[1, 1, 2, 1] + 
       10 I[1, 1, 3, 0]\Big) \nonumber\\
       &+ 
    4 \Big (5 I[0, 3, 1, 0] - 11 I[0, 3, 1, 1] + 9 I[0, 3, 1, 2] - 
        3 I[0, 3, 1, 3] - 10 I[0, 3, 2, 0] + 16 I[0, 3, 2, 1] 
        \nonumber\\
       &- 
        8 I[0, 3, 2, 2] + 5 I[0, 3, 3, 0] - 5 I[0, 3, 3, 1] + 
        15 I[1, 2, 1, 1] - 24 I[1, 2, 1, 2] + 9 I[1, 2, 1, 3] 
        \nonumber\\
       &- 
        30 I[1, 2, 2, 1] + 24 I[1, 2, 2, 2] + 
        10 I[1, 2, 3, 1]\Big)\Bigg\}  \nonumber\\
        %%%%%%%%%%%%%%%%%%%%%%%%%%%%%%%%%%%%%%%%%
      %  \end{align}
      %  \begin{align}
        &
        -e_u \Bigg\{3 m_ 0^2 \Big (2  m_s \big (I[0, 2, 1, 0] - 
          2 I[0, 2, 1, 1] + I[0, 2, 1, 2] - 2 I[0, 2, 2, 0] + 
          2 I[0, 2, 2, 1] + I[0, 2, 3, 0] 
          \nonumber\\
       &- 2 I[1, 1, 1, 0] - 
          4 I[1, 1, 1, 1] + 2 I[1, 1, 1, 2] - 4 I[1, 1, 2, 0] + 
          4 I[1, 1, 2, 1] + 2 I[1, 1, 3, 0]\big)
          \nonumber\\
    %       \end{align}
    %  \begin{align}
       &+ 
       m_c \big (9 I[0, 2, 1, 0] - 18 I[0, 2, 1, 1] + 
           9 I[0, 2, 1, 2] - 14 I[0, 2, 2, 0] + 14 I[0, 2, 2, 1] + 
           5 I[0, 2, 3, 0] 
           \nonumber\\
       &- 10 I[1, 1, 1, 0] - 28 I[1, 1, 1, 1] + 
           18 I[1, 1, 1, 2] - 20 I[1, 1, 2, 0] + 28 I[1, 1, 2, 1] + 
           10 I[1, 1, 3, 0]\big)\Big) 
           \nonumber\\
     %      \end{align}
     % \begin{align}
       &+ 
    4 m_c \Big (5 I[0, 3, 1, 0] - 19 I[0, 3, 1, 1] + 
       21 I[0, 3, 1, 2] - 7 I[0, 3, 1, 3] - 10 I[0, 3, 2, 0] + 
       24 I[0, 3, 2, 1] 
       \nonumber\\
       &- 12 I[0, 3, 2, 2] + 5 I[0, 3, 3, 0] - 
       5 I[0, 3, 3, 1] + 15 I[1, 2, 1, 1] - 36 I[1, 2, 1, 2] + 
       21 I[1, 2, 1, 3] 
       \nonumber \\      
      %\end{align}
     % \begin{align}
       & - 30 I[1, 2, 2, 1]+ 36 I[1, 2, 2, 2] + 
       15 I[1, 2, 3, 1]\Big) + 
    8  m_s \Big (I[0, 3, 1, 0] - 3 I[0, 3, 1, 1] + 3 I[0, 3, 1, 2] - 
        I[0, 3, 1, 3]\nonumber\\
              & - 2 I[0, 3, 2, 0] 
        + 4 I[0, 3, 2, 1] - 
        2 I[0, 3, 2, 2] + I[0, 3, 3, 0] - I[0, 3, 3, 1] + 
        3 I[1, 2, 1, 1] - 6 I[1, 2, 1, 2] + 3 I[1, 2, 1, 3]
        \nonumber\\
       & - 
        6 I[1, 2, 2, 1] + 6 I[1, 2, 2, 2] + 
        3 I[1, 2, 3, 1]\Big)\Bigg\} \Bigg]\nonumber\\
        %%%%%%%%%%%%%%%%%%%%%%%%%%%%%%%%%%%%%%%%%%%%%%%%%%%%%%%%%%%%%%%%%%%%
     %   \end{align}
      % \begin{align}
        &+\frac{m_c P_1 P_3}{679477248 \pi^5}\Bigg[2 e_c \Bigg \{m_ 0^2 \Big (6 m_c \big (5 I[0, 2, 1, 0] - 
          4 I[0, 2, 1, 1] - I[0, 2, 1, 2] - 5 I[0, 2, 2, 0] - 
          I[0, 2, 2, 1] - 2 I[1, 1, 1, 1]
          \nonumber\\
              &+ 2 I[1, 1, 1, 2] + 
          2 I[1, 1, 2, 1]\big) + 
       m_s \big (I[0, 2, 1, 0] - 2 I[0, 2, 1, 1] + I[0, 2, 1, 2] - 
           2 I[0, 2, 2, 0] + 2 I[0, 2, 2, 1] + I[0, 2, 3, 0] 
           \nonumber\\
              &- 
           2 I[1, 1, 1, 0] - 4 I[1, 1, 1, 1] + 2 I[1, 1, 1, 2] - 
           4 I[1, 1, 2, 0] + 4 I[1, 1, 2, 1] + 
           2 I[1, 1, 3, 0]\big)\Big) + 
    2 m_c \Big (3 I[0, 3, 1, 0]
    \nonumber\\
              &- 26 I[0, 3, 1, 1] + 
       21 I[0, 3, 1, 2] + 2 I[0, 3, 1, 3] - 6 I[0, 3, 2, 0] + 
       26 I[0, 3, 2, 1] + 2 I[0, 3, 2, 2] + 3 I[0, 3, 3, 0] + 
       6 I[1, 2, 1, 2] 
       \nonumber\\
              &- 6 I[1, 2, 1, 3] - 6 I[1, 2, 2, 2]\Big) + 
    2 m_s \Big (31 I[0, 3, 1, 0] - 63 I[0, 3, 1, 1] + 
        33 I[0, 3, 1, 2] - I[0, 3, 1, 3] - 62 I[0, 3, 2, 0] 
        \nonumber\\
              &+ 
        64 I[0, 3, 2, 1] - 2 I[0, 3, 2, 2] + 31 I[0, 3, 3, 0] - 
        I[0, 3, 3, 1] + 3 I[1, 2, 1, 1] 
        \nonumber\\
              &- 6 I[1, 2, 1, 2] + 
        3 I[1, 2, 1, 3] - 6 I[1, 2, 2, 1] + 6 I[1, 2, 2, 2] + 
        3 I[1, 2, 3, 1]\Big)\Bigg\}
        \nonumber\\
       & - (e_d + 
     e_u) \Bigg \{m_ 0^2 \Big (-2 m_s \big(I[0, 2, 1, 0] - 
          2 I[0, 2, 1, 1] + I[0, 2, 1, 2] - 2 I[0, 2, 2, 0] + 
          2 I[0, 2, 2, 1] + I[0, 2, 3, 0] 
          \nonumber\\
              &- 
          2I[1, 1, 1, 0] - 4 I[1, 1, 1, 1] + 2I[1, 1, 1, 2] - 
             4 I[1, 1, 2, 0] + 4 I[1, 1, 2, 1] +2 I[1, 1, 3, 0]\big) + 
       3 m_c \big(5 I[0, 2, 1, 0] 
       \nonumber\\
              &- 10 I[0, 2, 1, 1] 
       + 5 I[0, 2, 1, 2] - 
           6 I[0, 2, 2, 0] + 6 I[0, 2, 2, 1] + I[0, 2, 3, 0] - 
           2 (I[1, 1, 1, 0] - 6 I[1, 1, 1, 1] + 5 I[1, 1, 1, 2] \nonumber\\
              &- 
              2 I[1, 1, 2, 0] + 6 I[1, 1, 2, 1] + 
              I[1, 1, 3, 0])\big)\Big) + 
    m_s \Big (-I[0, 3, 1, 0] + 3 I[0, 3, 1, 1] - 3 I[0, 3, 1, 2] + 
       I[0, 3, 1, 3] 
       \nonumber\\
              &- I[0, 3, 2, 0] + 2 I[0, 3, 2, 1] - 
       I[0, 3, 2, 2] + 5 I[0, 3, 3, 0] - 5 I[0, 3, 3, 1] - 
       3 I[0, 3, 4, 0] - 15 I[1, 2, 1, 1] + 3 I[1, 2, 1, 2] 
       \nonumber\\
              &+ 
       3 I[1, 2, 1, 3] + 30 I[1, 2, 2, 1] - 3 I[1, 2, 2, 2] - 
       15 I[1, 2, 3, 1]\Big) + 
    4 m_c \Big (I[0, 3, 1, 0] - 7 I[0, 3, 1, 1] + 9 I[0, 3, 1, 2] 
    \nonumber\\
              &- 
        3 I[0, 3, 1, 3] - 2 I[0, 3, 2, 0] + 8 I[0, 3, 2, 1] - 
        4 I[0, 3, 2, 2] + I[0, 3, 3, 0] - I[0, 3, 3, 1] + 
        3 I[1, 2, 1, 1] - 12 I[1, 2, 1, 2] 
        \nonumber\\
              &+ 9 I[1, 2, 1, 3] - 
        6 I[1, 2, 2, 1] + 12 I[1, 2, 2, 2] + 
        3 I[1, 2, 3, 1]\Big)\Bigg\} \Bigg]\nonumber%\\
        \end{align}
\begin{align}
        %%%%%%%%%%%%%%%%%%%%%%%%%%%%%%%%%%%%%%%%%%%%%%%%%%%%
               & -\frac {5  e_c m_c P_ 2 P_ 3} {2359296 \pi^3}\Bigg[
   4 m_c m_s \Big (I[0, 3, 1, 0] - 2 I[0, 3, 1, 1] + I[0, 3, 1, 2] - 
       2 I[0, 3, 2, 0] + 2 I[0, 3, 2, 1] + I[0, 3, 3, 0]\Big)
        \nonumber\\
              &+ 
    3 \Big (-I[0, 4, 1, 0] + 3 I[0, 4, 1, 1] - 3 I[0, 4, 1, 2] + 
        I[0, 4, 1, 3] + 3 I[0, 4, 2, 0] - 6 I[0, 4, 2, 1] + 
        3 I[0, 4, 2, 2] - 3 I[0, 4, 3, 0] 
         \nonumber\\
              &+ 3 I[0, 4, 3, 1] + 
        I[0, 4, 4, 0]\Big)\Bigg]\nonumber\\
           & +\frac {m_c P_ 1} {27179089920 \pi^7}\Bigg[-e_c\Big (-108 I[0, 5, 1, 
         1] - 498 I[0, 5, 1, 2] + 673 I[0, 5, 1, 3] - 
       284 I[0, 5, 1, 4] + I[0, 5, 1, 5] 
       \nonumber\\
              &- 324 I[0, 5, 2, 1] + 
       996 I[0, 5, 2, 2] - 674 I[0, 5, 2, 3] + 2 I[0, 5, 2, 4] + 
       324 I[0, 5, 3, 1] - 498 I[0, 5, 3, 2] + I[0, 5, 3, 3]
       \nonumber\\
              &- 
       108 I[0, 5, 4, 1] + 
       10 m_c m_s \big (9 I[0, 4, 1, 1] - 48 I[0, 4, 1, 2] + 
          37 I[0, 4, 1, 3] + 2 I[0, 4, 1, 4] - 18 I[0, 4, 2, 1] + 
          48 I[0, 4, 2, 2] 
          \nonumber\\
              &+ 2 I[0, 4, 2, 3] + 9 I[0, 4, 3, 1] + 
          8 I[1, 3, 1, 3] - 8 I[1, 3, 1, 4] - 8 I[1, 3, 2, 3]\big) - 
       5 (I[1, 4, 1, 3] - 2 I[1, 4, 1, 4] + I[1, 4, 1, 5] 
       \nonumber\\
              &- 
           2 I[1, 4, 2, 3] + 2 I[1, 4, 2, 4] + I[1, 4, 3, 3])\Big) 
           \nonumber\\
    %    \end{align}
     %   \begin{align}
                %%%%%%%%%%%%%%%%%%%%%%%%%%%%%%%%%%%%%%%%%%%%%%%%%%%%%%%%%
        %%%%%%%%%%%%%%%%%%%%%%%%%%%%%%%%%%%%%%%%%%%%%%%%%%%%%%%%
       &+ 
    e_s \Big (90 I[0, 5, 1, 1] - 315 I[0, 5, 1, 2] + 
       419 I[0, 5, 1, 3] 
       - 235 I[0, 5, 1, 4] + 47 I[0, 5, 1, 5] - 
       270 I[0, 5, 2, 1]
       \nonumber\\
              & + 687 I[0, 5, 2, 2] - 556 I[0, 5, 2, 3] + 
       139 I[0, 5, 2, 4] + 270 I[0, 5, 3, 1] - 411 I[0, 5, 3, 2] 
       \nonumber\\
              &+ 
       137 I[0, 5, 3, 3] - 
       5 \big (18 I[0, 5, 4, 1] - 9 I[0, 5, 4, 2] - 
           45 I[1, 4, 1, 2] + 137 I[1, 4, 1, 3] - 139 I[1, 4, 1, 4] + 
           47 I[1, 4, 1, 5] 
           \nonumber\\
              &+ 135 I[1, 4, 2, 2] - 274 I[1, 4, 2, 3] + 
           139 I[1, 4, 2, 4] - 135 I[1, 4, 3, 2] + 
           137 I[1, 4, 3, 3] - 45 I[1, 4, 4, 2]\big)\Big)\nonumber\\
     %      \end{align}
      %     \begin{align}
           &+ 
    e_d \Big (54 I[0, 5, 1, 1] - 183 I[0, 5, 1, 2] + 
       229 I[0, 5, 1, 3] - 125 I[0, 5, 1, 4] + 25 I[0, 5, 1, 5] - 
       162 I[0, 5, 2, 1] + 393 I[0, 5, 2, 2] 
       \nonumber\\
              &- 308 I[0, 5, 2, 3] + 
       77 I[0, 5, 2, 4] + 162 I[0, 5, 3, 1] - 237 I[0, 5, 3, 2] + 
       79 I[0, 5, 3, 3] - 54 I[0, 5, 4, 1] + 27 I[0, 5, 4, 2] 
       \nonumber\\
              &+ 
       5 m_c m_s \big (6 I[0, 4, 1, 1] - 27 I[0, 4, 1, 2] + 
          28 I[0, 4, 1, 3] - 7 I[0, 4, 1, 4] - 12 I[0, 4, 2, 1] + 
          30 I[0, 4, 2, 2] - 10 I[0, 4, 2, 3]
          \nonumber\\
              &+ 6 I[0, 4, 3, 1] - 
          3 I[0, 4, 3, 2] + 12 I[1, 3, 1, 2] - 40 I[1, 3, 1, 3] + 
          28 I[1, 3, 1, 4] - 24 I[1, 3, 2, 2] + 40 I[1, 3, 2, 3] 
          \nonumber\\
              &+ 
          12 I[1, 3, 3, 2]\big) + 
       5 \big (27 I[1, 4, 1, 2] - 79 I[1, 4, 1, 3] + 
           77 I[1, 4, 1, 4] - 25 I[1, 4, 1, 5] - 81 I[1, 4, 2, 2] + 
           158 I[1, 4, 2, 3] 
           \nonumber\\
              &- 77 I[1, 4, 2, 4] + 81 I[1, 4, 3, 2] - 
           79 I[1, 4, 3, 3] - 27 I[1, 4, 4, 2]\big)\Big)\nonumber\\
              & + 
    e_u \Big (54 I[0, 5, 1, 1] - (189 - 6 I) I[0, 5, 1, 2] + 
        229 I[0, 5, 1, 3] - 125 I[0, 5, 1, 4] + 25 I[0, 5, 1, 5] - 
        162 I[0, 5, 2, 1] 
        \nonumber\\
              &+ 393 I[0, 5, 2, 2] - 308 I[0, 5, 2, 3] + 
        77 I[0, 5, 2, 4] + 162 I[0, 5, 3, 1] - 237 I[0, 5, 3, 2] + 
        79 I[0, 5, 3, 3] - 54 I[0, 5, 4, 1] 
        \nonumber\\
              &+ 27 I[0, 5, 4, 2] + 
        5 m_c m_s \big (6 I[0, 4, 1, 1] - 27 I[0, 4, 1, 2] + 
           28 I[0, 4, 1, 3] - 7 I[0, 4, 1, 4] - 12 I[0, 4, 2, 1] + 
           30 I[0, 4, 2, 2] 
           \nonumber\\
              &- 10 I[0, 4, 2, 3] + 6 I[0, 4, 3, 1] - 
           3 I[0, 4, 3, 2] + 12 I[1, 3, 1, 2] - 40 I[1, 3, 1, 3] + 
           28 I[1, 3, 1, 4] - 24 I[1, 3, 2, 2] 
           \nonumber\\
              &+ 40 I[1, 3, 2, 3] + 
           12 I[1, 3, 3, 2]\big) + 
        5 \big (27 I[1, 4, 1, 2] - 79 I[1, 4, 1, 3] + 
            77 I[1, 4, 1, 4] - 25 I[1, 4, 1, 5] - 81 I[1, 4, 2, 2] 
            \nonumber\\
              &+ 
            158 I[1, 4, 2, 3] - 77 I[1, 4, 2, 4] + 81 I[1, 4, 3, 2] - 
            79 I[1, 4, 3, 3] + 27 I[1, 4, 4, 2]\big)\Big)\Bigg]\nonumber\\
            %%%%%%%%%%%%%%%%%%%%%%%%%%%%%%%%%%%%%%%%%%%%%%%%
            %%%%%%%%%%%%%%%%%%%%%%%%%%%%%%%%%%%%%%%%%%%%%%%%%%%%%
              &-  \frac {e_c m_c P_ 2} {37748736 \pi^5}\Bigg[
   2 m_c \Bigg (15 m_ 0^2 \Big (I[0, 4, 1, 1] - 2 I[0, 4, 1, 2] + 
          I[0, 4, 1, 3] - 2 I[0, 4, 2, 1] + 2 I[0, 4, 2, 2] + 
          I[0, 4, 3, 1]\Big) 
          \nonumber\\
              &+ 
       6 \big (I[0, 5, 1, 2] - 2 I[0, 5, 1, 3] + I[0, 5, 1, 4] - 
           2 I[0, 5, 2, 2] + 2 I[0, 5, 2, 3] + 
           I[0, 5, 3, 2]\big)\Bigg) - 
    3 \Bigg (15 m_ 0^2 m_s \Big (I[0, 4, 1, 0] 
    \nonumber\\
              &- 3 I[0, 4, 1, 1] + 
           3 I[0, 4, 1, 2] - I[0, 4, 1, 3] - 
           3 (I[0, 4, 2, 0] - 2 I[0, 4, 2, 1] + I[0, 4, 2, 2] - 
              I[0, 4, 3, 0] + I[0, 4, 3, 1])
              \nonumber\\
              &- I[0, 4, 4, 0]\Big) + 
        12 m_s \Big (-I[0, 5, 1, 1] + 3 I[0, 5, 1, 2] - 
            3 I[0, 5, 1, 3] + I[0, 5, 1, 4] + 
            3 (I[0, 5, 2, 1] - 2 I[0, 5, 2, 2] 
            \nonumber\\
              &+ I[0, 5, 2, 3] - 
               I[0, 5, 3, 1] + I[0, 5, 3, 2]) + 
            I[0, 5, 4, 1]\Big)\Bigg)\Bigg]\nonumber%\\
            %%%%%%%%%%%%%%%%%%%%%%%%%%%%%%%%%%%%%%%%%%%%%%%%%%%
           \end{align}
       \begin{align}
 &-\frac {e_c m_c P_ 3} {125829120 \pi^5}  \Bigg[
   5 m_0^2 \Bigg (2 m_c \Big (I[0, 4, 1, 1] - 2 I[0, 4, 1, 2] + 
          I[0, 4, 1, 3] - 2 I[0, 4, 2, 1] + 2 I[0, 4, 2, 2] + 
          I[0, 4, 3, 1]\Big) 
          \nonumber\\
              &+ 
       11 m_s \Big (I[0, 4, 1, 0] - 3 I[0, 4, 1, 1] + 
           3 I[0, 4, 1, 2] - I[0, 4, 1, 3] - 
           3 (I[0, 4, 2, 0] - 2 I[0, 4, 2, 1] + I[0, 4, 2, 2] - 
              I[0, 4, 3, 0] 
              \nonumber\\
              &+ I[0, 4, 3, 1]) - 
           I[0, 4, 4, 0]\Big)\Bigg) - 
    4 m_c \Big (I[0, 5, 1, 2] - 2 I[0, 5, 1, 3] + I[0, 5, 1, 4] - 
       2 I[0, 5, 2, 2] + 2 I[0, 5, 2, 3] 
       \nonumber\\
       %  \end{align}
       % \begin{align}
              &+ I[0, 5, 3, 2]\Big) - 
    66 m_s \Big (I[0, 5, 1, 1] - 3 I[0, 5, 1, 2] + 3 I[0, 5, 1, 3] - 
        I[0, 5, 1, 4] - 
        3 (I[0, 5, 2, 1] - 2 I[0, 5, 2, 2] 
        + I[0, 5, 2, 3]\nonumber\\
              & - 
           I[0, 5, 3, 1] 
           + I[0, 5, 3, 2]) - I[0, 5, 4, 1]\Big)\Bigg]  \nonumber\\
           %%%%%%%%%%%%%%%%%%%%%%%%%%%%%%%%%%%%%%%%%%%%%%%%%%%%%%%%%%%%%%%%%%%%%%%
                      &
           -\frac {e_c m_c^2 m_s} {754974720 \pi^7}\Bigg[
   I[0, 6, 1, 3] - 2 I[0, 6, 1, 4] + I[0, 6, 1, 5] - 
    2 I[0, 6, 2, 3] + 2 I[0, 6, 2, 4] + I[0, 6, 3, 3]\Bigg]
    %%%%%%%%%%%%
    \nonumber\\
      &
    +\frac {11  e_c m_c} {3523215360 \pi^7} \Bigg[
   I[0, 7, 1, 3] - 3 I[0, 7, 1, 4] + 3 I[0, 7, 1, 5] - 
    I[0, 7, 1, 6] - 3 I[0, 7, 2, 3] + 6 I[0, 7, 2, 4] - 
    3 I[0, 7, 2, 5]
    \nonumber\\
    &+ 3 I[0, 7, 3, 3] - 3 I[0, 7, 3, 4] - 
    I[0, 7, 4, 3]\Bigg]\nonumber\\
   %         \end{align}
    %   \begin{align}
                     %%%%%%%%%%%%%%%%%%%%%%%%%%%%%%%%%%%%%%%%%%%%%%%%%%%%%%%%%%%%
           %%%%%%%%%%%%%%%%%%%%%%%%%%%%%%%%%%%%%%%%%%%%%%%%%%%%%%%%%%
        %     \end{align}
        %\begin{align}
 %%%%%%%%%%%%%%%%%%%%%%%%%%%%%%%%%%%%%%%%%%%%%%%%% 
 &+\frac {f_ {3\gamma} m_c P_ 1} {86973087744 \pi^5}\Bigg[
   64 \Bigg (ed \Big (-4 m_c m_s \big (I[0, 3, 1, 0] - 
             7 I[0, 3, 1, 1] + 9 I[0, 3, 1, 2] - 3 I[0, 3, 1, 3] - 
             2 I[0, 3, 2, 0] 
             \nonumber\\
    &+ 8 I[0, 3, 2, 1] - 4 I[0, 3, 2, 2] + 
             I[0, 3, 3, 0] - I[0, 3, 3, 1]\big) + 9 I[0, 4, 1, 0] - 
          38 I[0, 4, 1, 1] + 59 I[0, 4, 1, 2] 
          \nonumber\\
    &- 40 I[0, 4, 1, 3] + 
          10 I[0, 4, 1, 4] - 27 I[0, 4, 2, 0] + 85 I[0, 4, 2, 1] - 
          87 I[0, 4, 2, 2] + 29 I[0, 4, 2, 3] + 27 I[0, 4, 3, 0] 
          \nonumber\\
 %\end{align}
 %\begin{align}
& - 
          56 I[0, 4, 3, 1] + 28 I[0, 4, 3, 2] - 9 I[0, 4, 4, 0] + 
          9 I[0, 4, 4, 1]\Big) + 
       e_u \Big (-4 m_c m_s \big (I[0, 3, 1, 0] - 7 I[0, 3, 1, 1] + 
             9 I[0, 3, 1, 2] 
             \nonumber\\
    &- 3 I[0, 3, 1, 3] - 2 I[0, 3, 2, 0] + 
             8 I[0, 3, 2, 1] - 4 I[0, 3, 2, 2] + I[0, 3, 3, 0] - 
             I[0, 3, 3, 1]\big) + 9 I[0, 4, 1, 0] - 
          38 I[0, 4, 1, 1] 
          \nonumber\\
    &+ 59 I[0, 4, 1, 2] - 40 I[0, 4, 1, 3] + 
          10 I[0, 4, 1, 4] - 27 I[0, 4, 2, 0] + 85 I[0, 4, 2, 1] - 
          87 I[0, 4, 2, 2] + 29 I[0, 4, 2, 3] 
          \nonumber\\
    &+ 27 I[0, 4, 3, 0] - 
          56 I[0, 4, 3, 1] + 28 I[0, 4, 3, 2] - 9 I[0, 4, 4, 0] + 
          9 I[0, 4, 4, 1]\Big) - 
       e_s \Big (21 I[0, 4, 1, 0] - 82 I[0, 4, 1, 1] 
       \nonumber\\
    &+ 
           121 I[0, 4, 1, 2] - 80 I[0, 4, 1, 3] + 20 I[0, 4, 1, 4] - 
           63 I[0, 4, 2, 0] + 185 I[0, 4, 2, 1] - 183 I[0, 4, 2, 2] + 
           61 I[0, 4, 2, 3] 
           \nonumber\\
    &+ 63 I[0, 4, 3, 0] - 124 I[0, 4, 3, 1] + 
           62 I[0, 4, 3, 2] - 21 I[0, 4, 4, 0] + 
           21 I[0, 4, 4, 1]\Big)\Bigg) \varphi^a[u_ 0]
                      \nonumber\\
    &+
           I_ 2[\mathcal V] \Bigg (-45 (11 e_d + 24 e_s) I[0, 4, 3, 0] + 
   e_u \Big (736 m_c m_s I[0, 3, 2, 0] - 459 I[0, 4, 3, 0] - 
      30 I[0, 4, 4, 0]\Big) 
                 \nonumber\\
    &- 60 e_s I[0, 4, 4, 0]\Bigg)\Bigg]
 \nonumber\\
 %\end{align}
  %      \begin{align}
   %%%%%%%%%%%%%%%%%%%%%%%%%%%%%%%%%%%%
    &-\frac {m_c P_ 2 } {377487360 \pi^5}\Bigg[
   10 f_ {3\gamma}\pi^2 I_ 2[\mathcal V] \Big (20 e_s m_c (m_ 0^2 I[0,
             3, 3, 0] - I[0, 4, 3, 0]) + 
       e_u (10 m_ 0^2 m_c I[0, 3, 3, 0] - 10 m_c I[0, 4, 3, 0] 
       \nonumber\\
    &- 
          3 m_s I[0, 4, 4, 0]) + 
       2 e_d (5 m_ 0^2 m_c I[0, 3, 3, 0] - 5 m_c I[0, 4, 3, 0] + 
           9 m_s I[0, 4, 4, 0])\Big) + 
    3 I_ 4[\mathcal S] \Big (5 e_d m_c I[0, 5, 3, 0] 
    \nonumber\\
    &- 
        e_u m_c I[0, 5, 3, 0] + 21 e_d m_s I[0, 5, 4, 0] + 
        21 e_u m_s I[0, 5, 4, 0]\Big)\Bigg]\nonumber\\
                %%%%%%%%%%%%%%%%%%%%%%%%%%%%%%%%%%%%%%%%%%%%%%%%%%%%%%
  &+\frac {m_c P_ 3} {377487360 \pi^5}  \Bigg[
   5 f_ {3\gamma} \pi^2 I_ 2[\mathcal V] \Big (33 e_d m_s I[0, 4, 4, 
         0] + e_u (4 m_ 0^2 m_c I[0, 3, 3, 0] - 4 m_c I[0, 4, 3, 0] + 
           15 m_s I[0, 4, 4, 0])\Big) \nonumber\\
    &- 
    3 e_s m_c I_ 4[\mathcal S] I[0, 5, 3, 0]\Bigg]
    %%%%%%%%%%%%%%%%%%%%%%%%%%%%%%%%%%%%%%%%%%%%
    \nonumber\\
 %       \end{align}
 %       \begin{align}
    &+\frac {f_{3\gamma} m_c} {1006632960 \pi^5}\Bigg[
   I_ 2[\mathcal V] \Big (11 (e_d + e_s) I[0, 6, 4, 0] + 
       e_u (4 m_c m_s I[0, 5, 3, 0] + 5 I[0, 6, 4, 0])\Big)\Bigg],
       \label{appson}
 \end{align}
where $P_1 =\langle g_s^2 G^2\rangle$ is gluon condensate, $P_2 =\langle \bar q q \rangle$ stands for u/d-quark condensate and $P_3 =\langle \bar s s  \rangle$ denotes s-quark condensate . We should also remark that in Eq. (\ref{appson}), for simplicity we have only given the terms that give significant contributions to the numerical values of the magnetic moment and neglected to give many higher dimensional operators though they have been considered in the numerical calculations.
  The~$I[n,m,l,k]$, $I_1[\mathcal{F}]$, ~$I_2[\mathcal{F}]$, ~$I_3[\mathcal{F}]$, ~$I_4[\mathcal{F}]$, 
~$I_5[\mathcal{F}]$, and ~$I_6[\mathcal{F}]$ functions are
defined as:
\begin{align}
 I[n,m,l,k]&= \int_{4 m_Q^2}^{s_0} ds \int_{0}^1 dt \int_{0}^1 dw~ e^{-s/M^2}~
 s^n\,(s-4\,m_Q^2)^m\,t^l\,w^k,\nonumber\\
 %  \end{align}
 %\begin{align}
 I_1[\mathcal{F}]&=\int D_{\alpha_i} \int_0^1 dv~ \mathcal{F}(\alpha_{\bar q},\alpha_q,\alpha_g)
 \delta'(\alpha_ q +\bar v \alpha_g-u_0),\nonumber\\
  I_2[\mathcal{F}]&=\int D_{\alpha_i} \int_0^1 dv~ \mathcal{F}(\alpha_{\bar q},\alpha_q,\alpha_g)
 \delta'(\alpha_{\bar q}+ v \alpha_g-u_0),\nonumber\\
    I_3[\mathcal{F}]&=\int D_{\alpha_i} \int_0^1 dv~ \mathcal{F}(\alpha_{\bar q},\alpha_q,\alpha_g)
 \delta(\alpha_ q +\bar v \alpha_g-u_0),\nonumber\\
   I_4[\mathcal{F}]&=\int D_{\alpha_i} \int_0^1 dv~ \mathcal{F}(\alpha_{\bar q},\alpha_q,\alpha_g)
 \delta(\alpha_{\bar q}+ v \alpha_g-u_0),\nonumber\\
% \end{align}
% \begin{align}
   I_5[\mathcal{F}]&=\int_0^1 du~ \mathcal{F}(u)\delta'(u-u_0),\nonumber\\
 I_6[\mathcal{F}]&=\int_0^1 du~ \mathcal{F}(u),\nonumber
 \end{align}
 where $\mathcal{F}$ denotes the corresponding photon DAs.
 \end{widetext}

\bibliography{PcsMM}

\end{document}